\documentclass[11pt]{IEEEtran}
\usepackage{amsmath}
\usepackage[margin=2cm]{geometry}
\usepackage{amsmath}

\usepackage{tikz}
\usetikzlibrary{arrows}
\usetikzlibrary{decorations.markings}
\title{A Space-Time Approach for the Time-Domain Simulation in a Rotating Reference Frame}

\author{%
M. Klimek$^*$\thanks{* Graduate School of Computational Engineering, Technische Universit\"{a}t Darmstadt, 
Dolivo\-stra{\ss}e 15, D-64293, Darmstadt, Germany, e-mail:
\texttt{klimek@gsc.tu-darmstadt.de}, tel.: +49 6151 16-24391, fax: +49 6151 16-24404.}
\quad\and\quad
S. Kurz$^{*,\dagger}$
\quad\and\quad
S. Sch\"{o}ps$^{*,\dagger}$
\quad\and\quad
T. Weiland$^{\dagger}$\thanks{$\dagger$ Institut f\"{u}r Theorie Elektromagnetischer Felder, Technische Universit\"{a}t Darmstadt, 
Schlossgartenstra{\ss}e 8, D-64289, Darmstadt, Germany, e-mail:
\texttt{thomas.weiland@temf.tu-darmstadt.de}}
}

\begin{document}
\maketitle

\begin{abstract}
We approach the discretisation of Maxwell's equations directly in space-time
without making any non-relativistic assumptions
with the particular focus on simulations in rotating reference frames.
As a research example we study Sagnac's effect in a rotating ring resonator.
After the discretisation, we express the numerical scheme in a form resembling
3D FIT with leapfrog.
We compare the stability and convergence properties of two 4D approaches, namely FIT and FEM,
both using Whitney interpolation.
\end{abstract}

\section{INTRODUCTION}
Concerning Clifford's Geometric Algebra, we use the nomenclature and notation of \cite{GAForPhys},
which we also perceive as a comprehensive introduction to this mathematical formalism.
Maxwell's equations \cite[Eq.~(7.39)]{GAForPhys} read
\begin{align}
\nabla \wedge F = 0 \,, && \nabla \cdot G = J \,,
\label{ME4D}
\end{align}
where $F$ (or $G$) is the (dual) Faraday bivector, and $J$ the space-time current vector.

The bivectors $F$ and $G$ are related by constitutive material mapping $\xi$ defined implicitly
by
\begin{equation}
 G = \xi(F) \,.
\end{equation}
For linear, isotropic media $\xi$ is a space-time generalisation of the well known constitutive equations.
To derive an explicit formula for $\xi$, we start with
\begin{align}
\vec{D} = \varepsilon \vec{E} && \mbox{ and } && \vec{H} = \nu \vec{B} \,,
\end{align}
where $\varepsilon,\nu,\vec{E},\vec{H},\vec{D},\vec{B}$ are
electric permittivity, magnetic reluctance, electric and magnetic field strengths, and fluxes, respectively.
The resulting expression is given by
\begin{equation}
 \xi(F) = \frac{1}{2} \left[
  \left( \varepsilon + \frac{\nu}{c^2} \right) F -
  \left( \varepsilon - \frac{\nu}{c^2} \right) u F u
    \right] \,,
    \label{xi}
\end{equation}
with $u$ the four-velocity vector of the material,
and $c$ the speed of light in vacuum.

\section{MOTION AS MESH'S GEOMETRY}
%\subsection{Placement Map}
Motion of the system is modelled by specifying Lagrangian placement map $p_t(\vec{r}_\text{ref})$,
which gives a
space-time position (at time $t$) of a particle with initial position $\vec{r}_{\text{ref}}$.
The cylindrical coordinates of $\vec{r}_{\text{ref}}$ are denoted by $r, \varphi, z$.
For a system rotating with constant angular velocity $\Omega$ around $z$-axis
we use
\begin{align}
 p_t(\vec{r}_{\text{ref}}) =
 [t,r\cos(\theta),r\sin(\theta),z] %[\gamma_t, \gamma_x, \gamma_y, \gamma_z]^T 
 \,,
 \label{placement}
\end{align}
with $\theta = \varphi + \tanh\left(r\Omega/c\right) c t/r$.

%\subsection{4D Mesh Construction}
\begin{figure}[h]
\begin{tikzpicture}[scale=.21]
\node[inner sep=0pt]  at (0,0)
    {\includegraphics[
    width=.47\linewidth]{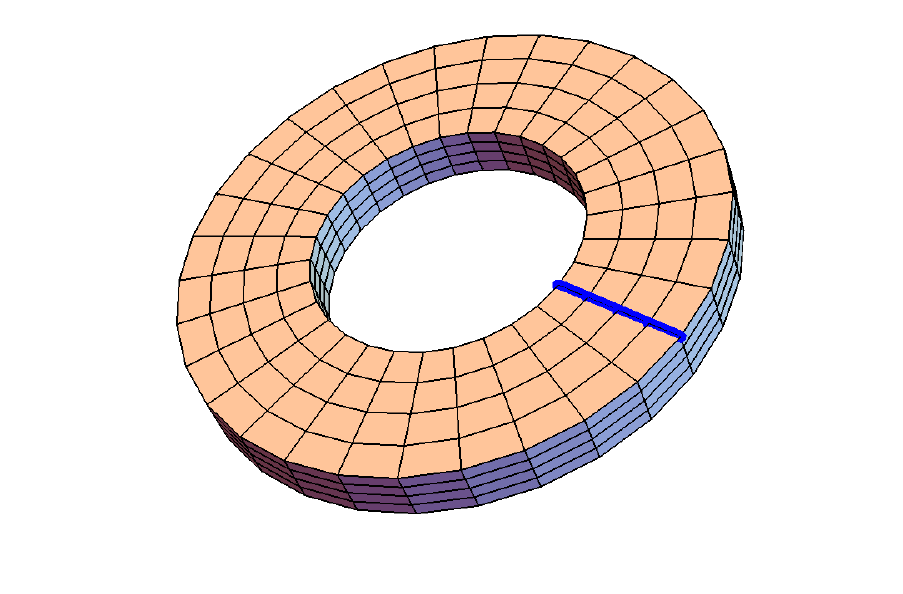}};
\node[inner sep=0pt]  at (20,0)
    {\includegraphics[width=.5\linewidth]{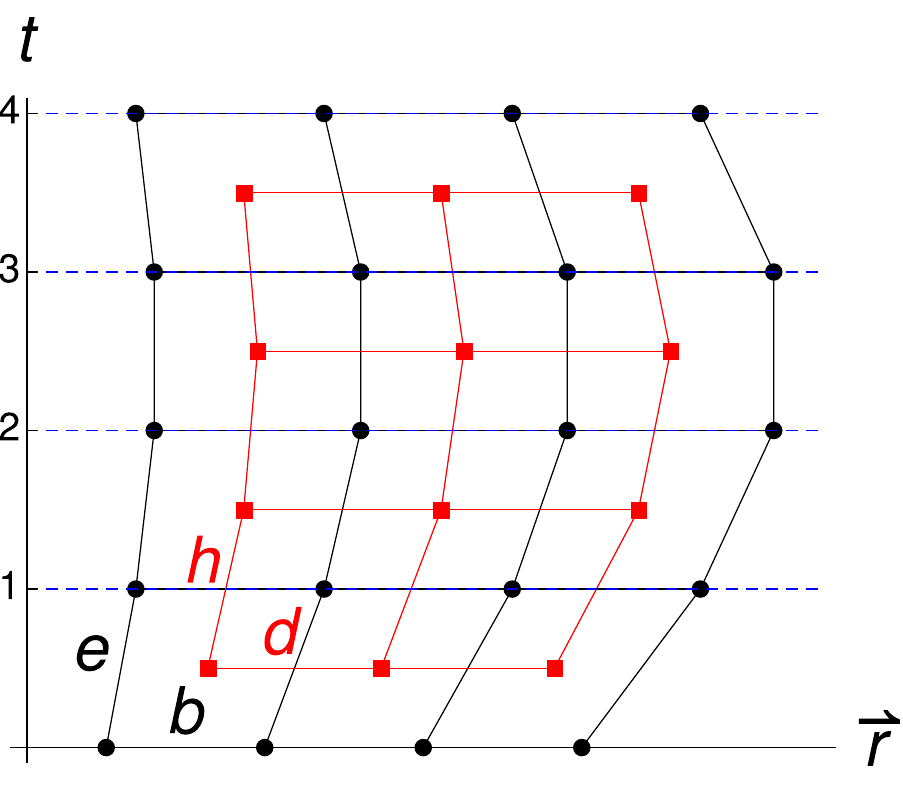}};
    \draw[
    decoration={markings,mark=at position 1 with {\arrow[scale=3]{>}}},
    postaction={decorate},
    shorten >=0.4pt
    ]
    (6.0,0.0) -- (10.0,-0.5);
    \draw[
    decoration={markings,mark=at position 1 with {\arrow[scale=3]{>}}},
    postaction={decorate},
    shorten >=0.4pt
    ]
    (6.3,1.0) -- (10.0,2.5);
        \draw[
    decoration={markings,mark=at position 1 with {\arrow[scale=3]{>}}},
    postaction={decorate},
    shorten >=0.4pt
    ]
    (6.0,2.0) -- (10.0,5.7);
    \draw[
    decoration={markings,mark=at position 1 with {\arrow[scale=3]{>}}},
    postaction={decorate},
    shorten >=0.4pt
    ]
    (5.8,-1) -- (10.0,-3.7);
    \draw[
    decoration={markings,mark=at position 1 with {\arrow[scale=3]{>}}},
    postaction={decorate},
    shorten >=0.4pt
    ]
    (5.5,-2) -- (10.0,-6.5);
\node[inner sep=0pt]  at (7,5.0) {$p_{t_i}
$};
\node[inner sep=0pt]  at (0.0,-6.5) {$\vec{r}_{\text{ref}}$};
\end{tikzpicture}
    \caption{
Sketch of a space-time mesh used in simulation.
The thick blue line in the left image is mapped to horizontal lines in the right figure.
}
    \label{fig:stmesh}
\end{figure}
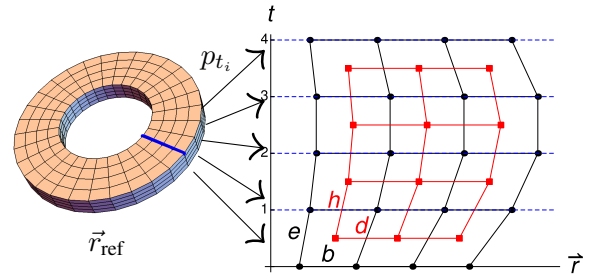

We extrude a 3D mesh by applying $p_{t_i}$ to nodes' positions for all time steps $t_i$
and connecting the 4D nodes stemming from the same 3D node;
see Fig.~\ref{fig:stmesh}.
By $K_n^i$ (respectively $\widetilde{K}_n^i$)
we denote the $i$-th $n$-dimensional element of the (barycentric dual) mesh.

\section{DISCRETIZATION}

\begin{figure}[h]
\begin{tikzpicture}[place/.style={circle,fill=black,inner sep= 0pt,minimum size = .3em}
, scale=1.7
]
  \coordinate [label={left:$m$}] (facet) at (0, 0);
  \coordinate [label={above:$l$}] (edge1) at (.15, .5);
  \coordinate (edge2) at (-.15, -.5);
  \coordinate (edge3) at (-.5, 0);
  \coordinate [label={right:$e^{n+1/2}_l$}] (e) at (1.5, 0);
  \coordinate [label={above:$b^n_m$}] (b) at (2, -.5);
  \coordinate [label={right:$K_2^j$}] (K) at (2.5,0);
  
  \coordinate (c1) at (1.5,-.5);
  \coordinate (c2) at (2.5,-.5);
  \coordinate (c3) at (2.5,+.5);
  \coordinate (c4) at (1.5,+.5);
  
  \node at (edge1) [style={circle,fill=black,inner sep= 0pt,minimum size = .6em}] {};
  \node at (edge2) [place] {};
  \node at (edge3) [place] {};
  \draw [red, line width = 2.0pt] (edge1) -- (edge2);
  \draw (edge1) -- (edge3) -- (edge2);
  
  \draw (c1) -- (c2) -- (c3) -- (c4) -- (c1);
  \node at (c1) [place] {};
  \node at (c2) [place] {};
  \node at (c3) [place] {};
  \node at (c4) [place] {};
  
  \path (edge1) edge [bend left,->,>=stealth',shorten >=3pt,semithick] node [above] {$p_{\left[ t_n,t_{n+1} \right]}$} (e);
  \path (facet) edge [bend right,->,>=stealth',shorten >=3pt,semithick] node [above] {$p_{t_n}$} (b);
  
  \coordinate (f0) at (-.7 , -.75);
  \coordinate (fx) at (.3 , -.75);
  \coordinate (fy) at (-.7, .25);
  
  \draw[thick,->] (f0) -- (fx) node[anchor=north east] {$x,y,z$};
  \draw[thick,->] (f0) -- (fy) node[anchor=north east] {$x,y,z$};
  
  \coordinate (f20) at (3 , -.75);
  \coordinate (f2x) at (2 , -.75);
  \coordinate (f2y) at (3, .25);
  
  \draw[thick,->] (f20) -- (f2x) node[anchor=north west] {$x,y,z$};
  \draw[thick,->] (f20) -- (f2y) node[anchor=north west] {$t$};
  
\end{tikzpicture}
\caption{
Illustration of the relation between indices in \eqref{DoFsSplit}.
The leftmost is the reference mesh. The $l$-th edge is depicted as a fat dot, and $m$-th 2D facet as a thick red line.
The rightmost is the space-time mesh.
2D facets are depicted as lines.
}
\label{fig:indices}
\end{figure}
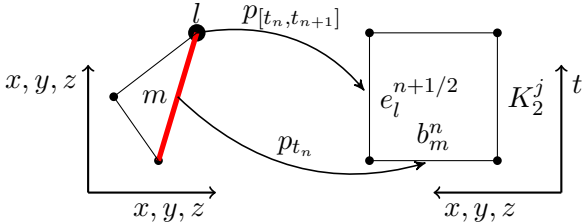

%\subsection{Maxwell's Equations}
Maxwell's equations \eqref{ME4D} are discretised by applying their integral form
to the primal/dual mesh pair, i.e.,
\begin{align}
  \oint\limits_{\partial K_3^i} (d^2x) \cdot F = 0 \,,
 &&
 \oint\limits_{\partial \widetilde{K}_3^i} ( d^2x ) \wedge G = \int\limits_{\widetilde{K}_3^i} (d^3x) \wedge J \,,
\end{align}
and introducing scalar DoFs (Degrees of Freedom) on the primal mesh
\begin{equation}
 f_j := \int\limits_{K_2^j} (d^2x) \cdot F \,,
% &&
% g_j := I^{-1} \int\limits_{\widetilde{K}_2^j} ( d^2x ) \wedge G \,,
% &&
% j_i := I^{-1}\int\limits_{\widetilde{K}_3^i} (d^3x) \wedge J \,,
 \label{DoFs}
\end{equation}
and analogously for the dual mesh.
Next, we rename and renumber the DoFs on the primal mesh as follows
\begin{equation}
  f_j %= \int\limits_{K_2^j} (d^2x) \cdot F 
  =: 
 \begin{cases}
  e^{n+1/2}_l & \mbox{if $K_2^j$ is timelike} \,,\\
  b^{n}_m & \mbox{if $K_2^j$ is spacelike} \,,
 \end{cases}
 \label{DoFsSplit}
\end{equation} 
where the relation between the indices $j$, $l$, $m$ and $n$ is as follows (see also Fig.~\ref{fig:indices}).
If the facet $K_2^j$ is timelike, 
it is the edge $l$ in the reference mesh $M^3$
extruded in time via the placement map $p_t$ such that $t \in \left[t_n,t_{n+1} \right]$.
Similarly, if the facet $K_2^j$ is spacelike,
then it is the image of a facet with index $m$ in $M^3$
under the map $p_t$ with $t=t_n$.
Analogously on the dual mesh, we split $g_j$ into $d^{n-1/2}$ and $h^{n}$.

\subsection{Material Equations}
A discrete equivalent of $\xi$ in \eqref{xi} is a material matrix $M_\xi$ relating
discrete equivalents of $F$ and $G$, i.e.,
\begin{equation}
 g_i = \sum_{j} \left[ M_\xi \right]_{ij} f_j \,.
\end{equation}
In both methods described below, we use $n$-vector valued Whitney interpolating functions
$N_i^n$ associated with $i$-th $n$-dimensional element of the mesh
to derive explicit expressions for $M_\xi$.

\subsubsection{4D FEM}
Applying energetic approach \cite{EnergeticApproach} also known as Galerkin Hodge star \cite{GalerkinHodge},
we arrive at the material matrix with entries
\begin{equation}
 \left[M_\xi^\text{FEM}\right]_{jk} := \int\limits_{K_4} N^2_j \cdot \xi(N^2_k) |d^4x| \,.
 \label{MMsFEM}
\end{equation}

\subsubsection{4D FIT}
The 4D FIT material matrix with Whitney interpolation written using Geometric Algebra reads
\begin{equation}
 \left[M_\xi^\text{FIT}\right]_{jk} := I^{-1} \widetilde{W}_{j} \wedge \bar{\xi}_j \left( N^2_k(x_j)\right) \,,
 \label{MMsFIT}
\end{equation}
with $\bar{\xi}_j$ is $\xi$ averaged over the $j$-th dual facet,
and $\widetilde{W}_{j}$ the bivector associated with it, describing its magnitude and orientation in space-time.
Although the collocation point $x_j$ may be chosen arbitrarily, 
we choose the barycenter of $K_2^j$ due to ease of calculation.

\subsubsection{Reduction to 3D Material Matrices and their Symmetrisation}
The material matrix $M_\xi$ is split into 3D material matrices $M_{\varepsilon/\nu,e/b}^\pm$ to resemble a 3D time marching scheme.
E.g., $M_{\varepsilon b}^+$ is the block of $M_\xi$ relating $d^{n+1/2}$ (role of $\varepsilon$)
to the future (role of $+$) magnetic $b$ DoF (role of $b$), i.e., $b^{n+1}$.

While $M_\xi^\text{FEM}$ is symmetric by construction \eqref{MMsFEM},
the FIT material matrix $M_\xi^\text{FIT}$ is symmetrised in order to avoid instabilities, as explained in \cite[especially Sec.~II]{RolfStability}.
The symmetrisation of $M_\xi^\text{FIT}$ translates to the following redefinitions,
where $M^T$ is the transpose of $M$,
\begin{align}
M^\text{FIT,sym}_{\nu b} &:= \frac{1}{2} \left( M^\text{FIT}_{\nu b} +\left[ M^\text{FIT}_{\nu b} \right]^T \right) \\
M^\text{FIT,sym}_{\varepsilon e} &:= \frac{1}{2} \left( M^\text{FIT}_{\varepsilon e} +\left[ M^\text{FIT}_{\varepsilon e} \right]^T \right) \\
M^{\pm \text{FIT,sym}}_{\varepsilon b} &:= \frac{1}{2} \left( M^{\pm \text{FIT}}_{\varepsilon b} + \left[ M^{\mp \text{FIT}}_{\nu e} \right]^T \right) \\
M^{\pm \text{FIT,sym}}_{\nu e} &:= \frac{1}{2} \left( M^{\pm \text{FIT}}_{\nu e} + \left[ M^{\mp \text{FIT}}_{\varepsilon b} \right]^T \right) \\
M^{\pm \text{FIT,sym}}_{\nu b} &:= \frac{1}{2} \left( M^{\pm \text{FIT}}_{\nu b} + \left[ M^{\mp \text{FIT}}_{\nu b} \right]^T \right)
\,.
\end{align}

\subsection{Resulting System of Linear Equations}
We have split space-time material matrix and DoFs into their 3D counterparts in a way, that the obtained
numerical scheme resembles 3D FIT with leapfrog, i.e.,
\begin{align}
h^n &= M_{\nu b}^-b^{n-1} + M_{\nu e}^-e^{n-1/2} + M_{\nu b} b^n + \nonumber \\
&+ M_{\nu e}^+e^{n+1/2} + M_{\nu b}^+b^{n+1} \nonumber \\
d^{n+1/2} &= d^{n-1/2} + \widetilde{C} h^n  \nonumber \\
e^{n+1/2} &= M_{\varepsilon e}^{-1} \left[ d^{n+1/2} - M_{\varepsilon b}^-b^{n} - M_{\varepsilon b}^+b^{n+1}  \right] \nonumber \\
b^{n+1} &= b^n + C e^{n+1/2} \,.
\label{resultingSystem}
\end{align}
We would like to note, that if $\Omega =0$ then all material matrices except $M_{\varepsilon e}$ and $M_{\nu b}$ vanish,
and on Cartesian grid the 4D FIT is naturally related to 3D FIT with leapfrog.

\section{STABILITY}
Starting from \eqref{resultingSystem} we derive the recursive formula
\begin{equation}
 \begin{bmatrix}
  b^{n+1} \\ e^{n+1/2} \\ b^{n} 
 \end{bmatrix}
 =
 U^n
 \begin{bmatrix}
  b^{1} \\ e^{1/2} \\ b^{0}
 \end{bmatrix} \,,
\end{equation}
where $U$ is the update matrix
\begin{equation}
 U := 
 \begin{bmatrix}
  1 + C M^{-1} \gamma & C M^{-1} \beta & C M^{-1} \alpha \\
  M^{-1} \gamma & M^{-1} \beta & M^{-1} \alpha \\
  1 & 0 & 0
 \end{bmatrix}
 \,,
 \label{updateMatrix}
\end{equation}
where $0$ and $1$ above are zero and identity matrices of proper dimensions and
\begin{align*}
 M :=& M_{\varepsilon e} - \widetilde{C} M_{\nu e}^+ - \widetilde{C} M_{\nu b}^+ C + M_{\varepsilon b}^+ C \\
 \alpha :=& M_{\varepsilon b}^- + \widetilde{C} M_{\nu b}^- \\
 \beta :=& M_{\varepsilon e} + \widetilde{C} M_{\nu e}^- \\
 \gamma :=& \widetilde{C} M_{\nu b} + \widetilde{C} M_{\nu b}^+ - M_{\varepsilon b}^-
 \,.
\end{align*}
The solution vector $[b^{n},e^{n-1/2}]$ will stay bounded if the modulus of all eigenvalues $\lambda$ of $U$ is less or equal to unity,
$|\lambda| \le 1$.
For both methods considered, the greatest $|\lambda|$ calculated on exemplary meshes used in simulations
is at most $1$.

As another test of stability,
we initialise the solver with randomly generated initial values and observe that
the norm of the solution does not grow in time.

\section{CONVERGENCE}
\begin{figure}[htpb]
\centerline{\includegraphics[width=\linewidth]{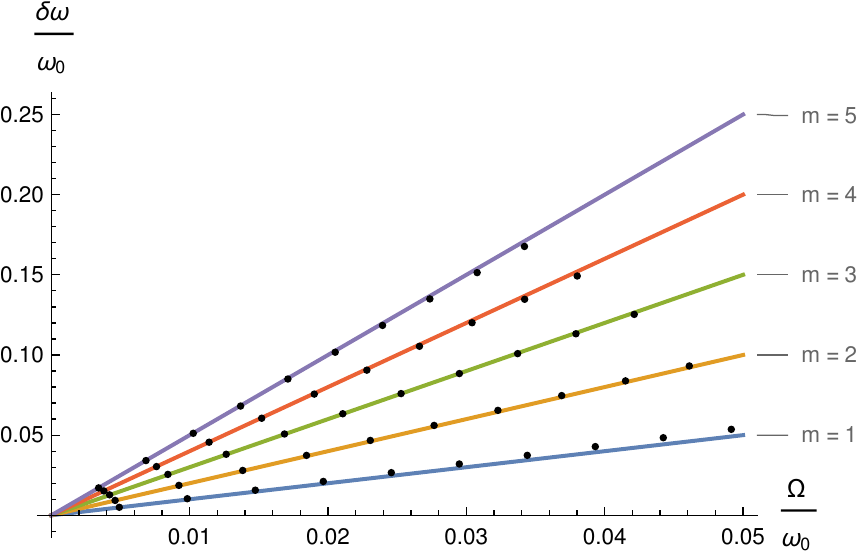}}
\caption{Comparison of analytical (solid lines) and numerical (points) frequency shifts in case of non-relativistic velocities.}
\label{fig:linearPlot}
\end{figure}

In order to verify the convergence of the scheme,
we simulate the first six, $m=0 \dots 5$, eigenmodes of the ring resonator structure studied in \cite[Section~IV]{Steinberg},
and compare the rotation induced frequency shift $\delta\omega^\text{sim}$ extracted from our numerical simulation
with the non-relativistic semi-analytic approximation,
\cite[Eq.~(4.5)]{Steinberg},
$\delta\omega^\text{anal} = m \Omega$.
As depicted in Fig.~\ref{fig:linearPlot} both approaches agree well
in a non-relativistic regime.
However,
in relativistic cases the relative difference
\begin{equation}
\eta := \frac{|\delta \omega^{\text{sim}} - \delta\omega^\text{anal}|}{|\delta\omega^\text{anal}|}
\label{relativeErr}
\end{equation}
between two approaches becomes significant, see Table~\ref{tab:FIT}.
We expect our solution to be correct as we have not made any non-relativistic assumptions
as opposed to \cite{Steinberg}.
\begin{table}
\scriptsize
\begin{equation*}
\begin{array}{|c|c|c|c|c|c|c|}
\hline  & \multicolumn{6}{c|}{v_{\text{max}}/c} \\
\hline m & <100 \%&99.63 \%&30.42 \%&3.14 \%&0.31 \%&0.03 \% \\ \hline
1& 95.7 \% & 58.1 \% & 2.2 \% & 0.0 \% & 0.2 \% & 3.2 \% \\ \hline
2& 95.8 \% & 58.5 \% & 2.8 \% & 0.9 \% & 0.9 \% & 1.0 \% \\ \hline
3& 95.8 \% & 59.2 \% & 2.6 \% & 2.0 \% & 3.9 \% & 3.4 \% \\ \hline
4& 96.0 \% & 60.6 \% & 4.6 \% & 1.3 \% & 1.3 \% & 2.4 \% \\ \hline
5& 96.2 \% & 62.2 \% & 5.0 \% & 8.2 \% & 8.2 \% & 2.6 \% \\ \hline
\end{array}
\end{equation*}
\caption{FIT case: the relative difference $\eta$
vs. mode number and rotation rate/velocity of the outer rim of the ring}
\label{tab:FIT}
\end{table}

\subsection{3D Wave Simulation without Rotation}
The requirements of the convergence proof in \cite{BossavitProof} are neither fulfilled by
$M_\xi^\text{FIT,sym}$ nor by $M_\xi^\text{FIT}$ (due to asymmetry).
Since this is the feature of the proposed method itself,
rather than its space-time extension, 
we focus now on 3D wave-simulation in the non-rotating ring resonator.
We investigate the convergence by comparing results obtained using 4D FIT material matrix $M_\xi^\text{FIT,sym}$
with the ones using $M_\xi^\text{FEM}$, for which the proof \cite{BossavitProof} holds.
We use temporal $L^2$ norm of a function of time $w: t \mapsto w(t)$
\begin{equation}
 \|w\|_2 := \sqrt{\int\limits_{t_\text{min}}^{t_\text{max}} dt \left[ w \left(t\right)\right]^2} \,,
\end{equation}
to define the relative difference
\begin{equation}
\text{diff}
% \Delta
 \left(E^z_1,E^z_2\right) := \frac{\|E^z_2(\vec{r}_\text{sample}) - E^z_1(\vec{r}_\text{sample}) \|_2}{\|E^z_1(\vec{r}_\text{sample}) \|_2}\,,
 \label{relativeDifference}
\end{equation}
where $E^z_{1}(\vec{r}_\text{sample})$ is the time signal at the sample point $\vec{r}_\text{sample}$,
obtained via interpolation of DoFs calculated using method and mesh indexed as $1$.

\begin{figure}[t]
\centering
\includegraphics[width=.45\linewidth]{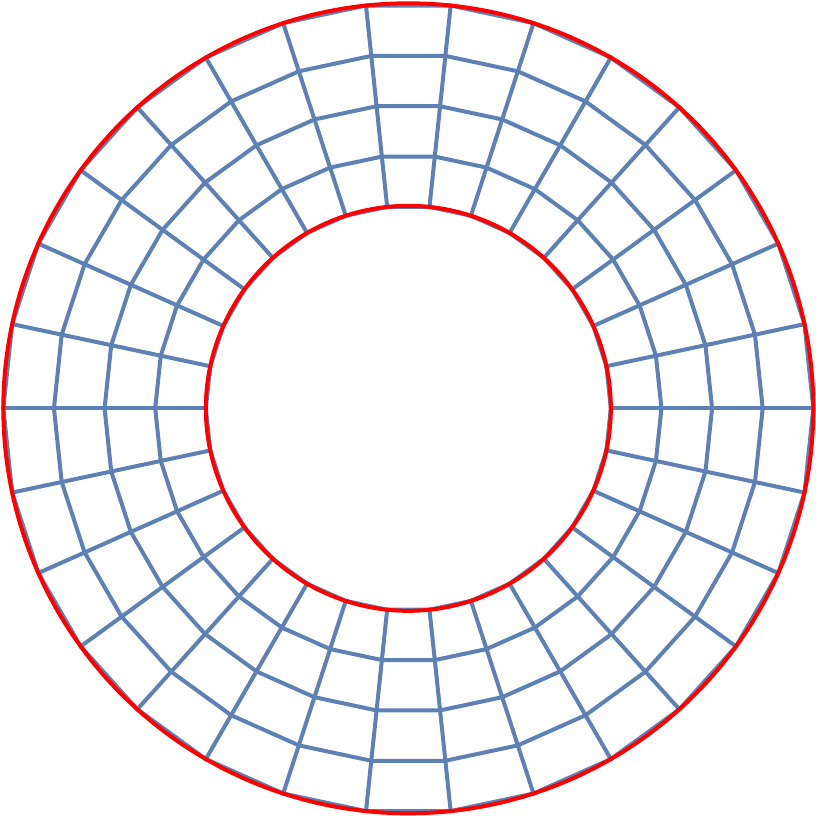}
\,
\includegraphics[width=.45\linewidth]{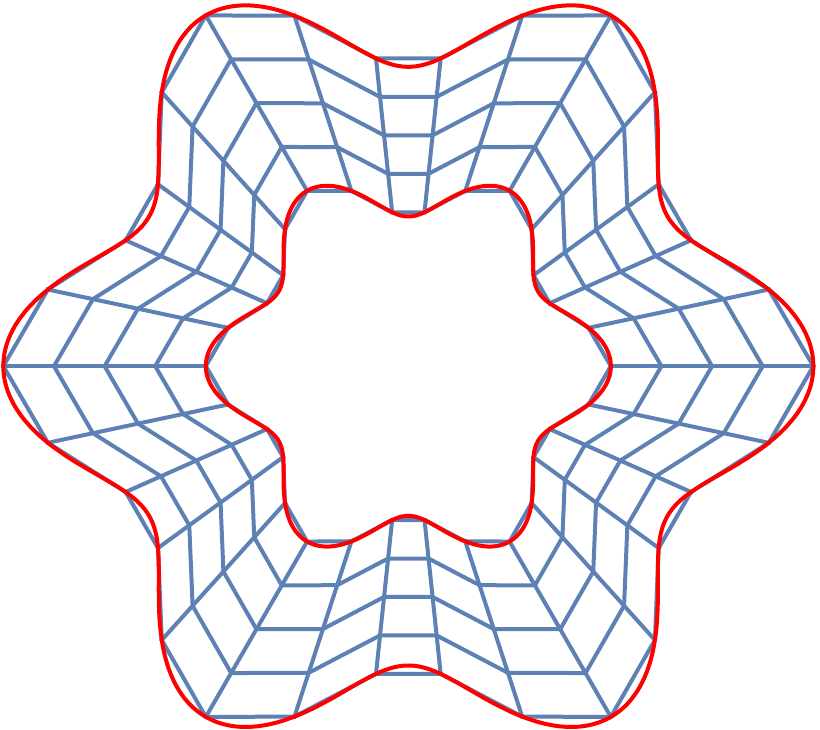}
\caption{Orthogonal (left) and non-orthogonal (right) mesh.}
\label{fig:meshes}
\end{figure}

\begin{figure}[h!]
\centerline{\includegraphics[width=\linewidth]{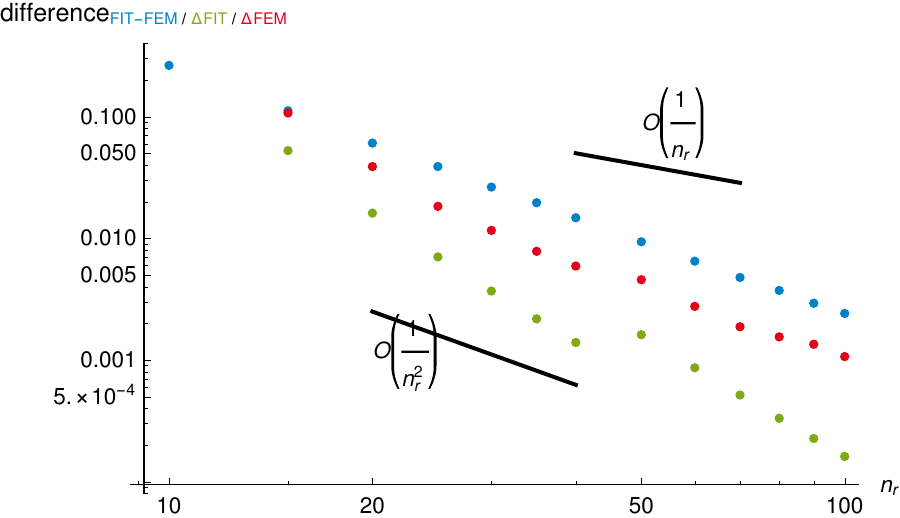}}
\centerline{\includegraphics[width=\linewidth]{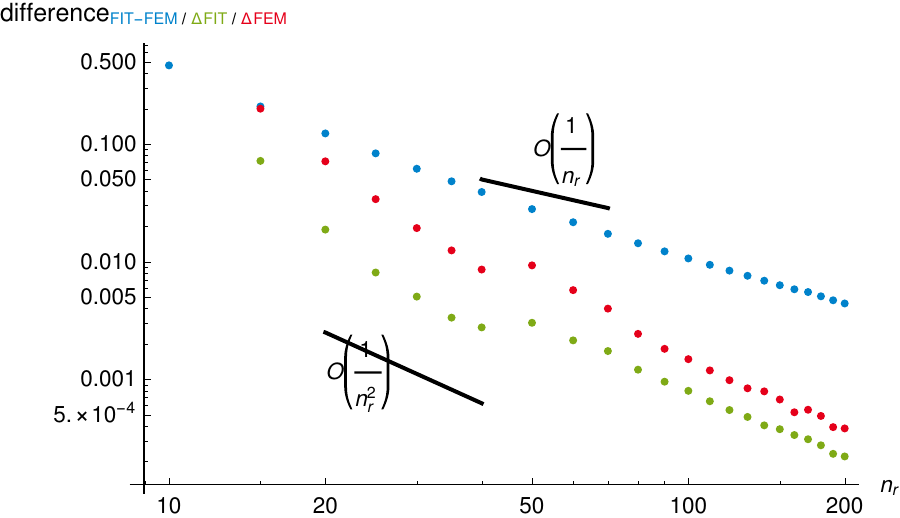}}
\caption{The relative difference \eqref{relativeDifference} between two successive FIT (green), FEM (red),
and FIT and FEM (blue) solutions on the mesh with $n_r$ nodes per wavelength.
Top: (Bottom:) results obtained for (non-)orthogonal mesh, left (right) in Fig.~\ref{fig:meshes}.}
\label{fig:convergence}
\end{figure}

The results for a (non-)orthogonal mesh, left (right) in Fig.~\ref{fig:meshes},
are depicted in top (bottom) of Fig.~\ref{fig:convergence}.

The proof \cite{BossavitProof} can be applied to FEM independently of orthogonality of the mesh,
therefore the convergence of FEM in both parts of Fig.~\ref{fig:convergence} is in accordance with that theory.
In case of orthogonal grid $M_\xi^\text{FIT}$ is diagonal, thus symmetric, and
the proof \cite{BossavitProof} can be applied in this particular situation.
Although, on non-orthogonal meshes the proof \cite{BossavitProof} cannot be repeated,
we observe that $M_\xi^\text{FIT,sym}$ gives a solution that converges
to the same solution as FEM (and with a similar rate of convergence).

\section{CONCLUSION}
We have applied space-time discretisation in two flavours, namely 4D FIT and FEM,
without making any non-relativistic approximations.
As a verification we recovered with good accuracy the rotation induced frequency shifts of the rotating ring resonator
predicted by an alternative non-relativistic approach.
The material matrix of our proposed extension of FIT is symmetrised to avoid instabilities.
With taking only the symmetric part of the material matrix,
the convergence is not guaranteed.
However, we investigated numerically that FIT converges to the same solution as FEM,
which is known to converge.

\section*{Acknowledgments}
The work of the first, second and third author is supported by 
the 'Excellence Initiative' of the German Federal and State Governments 
and the Graduate School of CE at Technische Universitaet Darmstadt.

\end{document}